\documentstyle[preprint,aps,prl]{revtex}
\draft
\begin{document}
\preprint{\today}
\title{Scaling behavior in the dynamics of a supercooled
Lennard-Jones mixture}
\author{Walter Kob}
\address{Institut f\"ur Physik, Johannes Gutenberg-Universit\"at,
Staudinger Weg 7, D-55099 Mainz, Germany}
\author{Hans C. Andersen}
\address{Department of Chemistry, Stanford University, Stanford,
California 94305}
\maketitle

\begin{abstract}
We present the results of a large scale molecular dynamics computer
simulation of a binary, supercooled Lennard-Jones fluid. At low
temperatures and intermediate times the time dependence of the
intermediate scattering function  is well described by a von Schweidler
law. The von Schweidler exponent is independent of temperature and
depends only weakly on the type of correlator.  For long times the
correlation functions show a Kohlrausch behavior with an exponent
$\beta$ that is independent of temperature.  This dynamical behavior is
in accordance with the mode-coupling theory of supercooled liquids.

\end{abstract}

\narrowtext
%\twocolumn


\par
In the last ten years there has been an impressive increase of our
understanding of the dynamics of supercooled liquids. In particular the
so-called mode-coupling theory (MCT), proposed by G\"otze and Sj\"ogren and,
independently, by Leutheusser, seems to offer a remarkable theoretical
insight into the dynamics of supercooled liquids\cite{bgsleuth84}.
However, despite the fact that certain experiments seem to be in
remarkable accordance with the theory\cite{cummins93}, the
applicability of the theory to supercooled liquids is still highly
controversial.\par

 In this paper we report some of the results of our efforts to
investigate the dynamics of a simple supercooled liquid in order to
find out whether the dynamics of this system is in accordance with the
predictions of MCT. A more extensive report will be given
elsewhere\cite{kob94a,kob94b}.  The reader is referred to
two review
articles by G\"otze and G\"otze and Sj\"ogren\cite{bibles} for a
discussion of the predictions of MCT.

 The model investigated is a binary mixture of Lennard-Jones particles
(800 particles of type A and 200 particles of type B).  Both kinds of
particles have the same mass. The parameters of the Lennard-Jones
potential were chosen as follows:  $\epsilon_{AA}=1.0$,
$\sigma_{AA}=1.0$, $\epsilon_{AB}=1.5$, $\sigma_{AB}=0.8$,
$\epsilon_{BB}=0.5$, and $\sigma_{BB}=0.88$.  The equations of motion
were integrated with the velocity form of the Verlet algorithm with a
time step of 0.01 and 0.02 at high and low temperatures respectively.
(Throughout the paper we use reduced time units, with the unit of time
being $(m\sigma_{AA}^2/48\epsilon_{AA})^{1/2}$, where $m$ is the mass of
a particle.)  The length of the
runs at the lowest temperature was $5\cdot10^{6}$ time steps, which
corresponds to a real time of about 10ns. More details on the
simulation will be given elsewhere\cite{kob94b}.\par

 In order to investigate the dynamical behavior of the system we
computed $F_{s}(q,t)$ and $F(q,t)$, the self and total intermediate
scattering function for wave vector $q$\cite{hansenmcdonald86}.  We
found that at low temperatures these correlation functions, when
plotted versus the logarithm of time, show a two-step relaxation
process with a well defined plateau at intermediate times, in agreement
with the predictions of MCT.  The width of the plateau is a strong
function of temperature, and at the lowest temperature it extends from
about
3 time
units to $10^{3}$ time units\cite{kob94a}.
In MCT, the approach to the plateau and the early stages of deviation
from it are referred to as the $\beta$-regime, whereas the entire
departure from the plateau, from the early stages to infinity, is
referred to as the $\alpha$-regime.
Even at the
lowest temperature all correlation functions that we measured decay to
zero within the time of our simulation.  This gives strong evidence that we
are able to equilibrate this system at all temperatures
investigated. Thus the results reported are all {\it equilibrium}
properties of the liquid.

 We defined the $\alpha$-relaxation time $\tau(T)$ as the time required
for
the correlation function to decay to $e^{-1}$ of its initial value.
Figure~\ref{fig1} shows $F_{s}(q,t)$ for the A particles versus the
rescaled time $t/\tau(T)$ for all temperatures investigated (see figure
caption). The wave vector $q$ is $q_{max}=7.25$, the location of the
maximum of the structure factor for the AA correlation. We recognize
that for low temperatures the curves follow a master curve that
extends
throughout the $\alpha-$relaxation regime. Thus this master curve extends over
about four orders of magnitude in time and is therefore quite remarkable.\par

 In the last part of the $\beta$-regime, which overlaps the
early part of the $\alpha$-regime, MCT predicts this master curve to
show a von Schweidler behavior, i.e. that the correlator $\phi(t)$ is
of the form $\phi(t)=f_{c}-A(t/\tau)^{b}$ with $A>0$. The constant $f_{c}$ is
called nonergodicity parameter and the positive exponent $b$ is called
the von Schweidler exponent. We made a fit to the master curve with
this functional form and the best fit obtained is included in the
figure as well. It is clear that the von Schweidler law describes the
relaxation behavior of the correlation function over about three
decades of rescaled time, which is remarkably long. For the value of
the von Schweidler exponent we obtained 0.49. In the later part of
the $\alpha$-regime, MCT predicts that the decay of the correlation
functions is well described by a Kohlrausch-Williams-Watt-law (KWW),
i.e. by $\phi(t)=A\exp(-(t/\tau(T))^{\beta})$. The best fit with this
functional form is included in the figure as well and we recognize that
this type of fit gives a good representation of the data. Note that the
value of the KWW exponent $\beta$ is 0.83 and therefore definitely
different from the value of the von Schweidler exponent $b$, as
predicted by MCT. Thus we
can conclude that the power-law observed for short rescaled time is not
just the short time expansion of the KWW function.

 MCT predicts that the von Schweidler exponent $b$ should be
independent of the value of the wave vector $q$ and also of the type
of correlator.  In a separate work \cite{kob94a} it was shown that for
the self intermediate scattering function of the A particles the value
of $b$ is almost independent of $q$ for values of $q$ between
$q_{max}=7.25$ and $q_{min}=9.61$, where $q_{min}$ is the location of
the first minimum in the structure factor for the AA correlation. Thus
this prediction of the theory seems to hold for the system under
investigation.\par

 Figure~\ref{fig2} shows the self intermediate scattering function
$F_s(q,t)$ for the B particles versus rescaled time $t/\tau(T)$. The
relaxation time $\tau(T)$ is defined here in the same way as for the
case of the A particles. The value of $q$ is $5.75$, the location of
the maximum of the structure factor for the BB correlation. As in the
case of the A particles we observe that at low temperatures the
correlator fall onto a master curve.  Also included in the figure is
the result of a fit with a von Schweidler law to this master curve in
the appropriate range of times (see above).  The von Schweidler
exponent is in this
case 0.47 and thus very close to the one found for the A particles.
Also in this case the long time behavior of the correlation functions
at low temperatures can be fitted well with a KWW function. The best
fit is included in the figure as well.\par

 Figure~\ref{fig3} shows the total intermediate scattering function
$F(q,t)$ for the A particles versus rescaled time $t/\tau(T)$. For $q$
we chose again $q_{max}$=7.25. Again we observe the presence of a
master curve that can be fitted well with a von Schweidler law. For
this correlator the von Schweidler exponent is 0.52 which is reasonably
close to the value we reported above for the values of the von
Schweidler exponent. Also for this correlation function the long time
behavior could be fitted well with a KWW function.\par

 In summary we can conclude, that the system investigated here seems to
exhibit a dynamical behavior that is in accordance with the
predictions of mode-coupling theory. In particular we find that the
correlation functions show a two-step relaxation behavior.
Furthermore, we find that in the late $\beta$-regime or early
$\alpha$-regime
 the
correlation functions investigated show a von Schweidler law.  The von
Schweidler exponent is only a weak function of the wave vector $q$ and
of the type of correlation function. Thus with respect to this the
predictions of MCT seem to hold. Also the observation that the later
stages of
$\alpha$-relaxation of the correlation functions can be fitted well by
a KWW function, with an exponent $\beta$ that is independent of $T$,
is in accordance with the theory.

Acknowledgments: Part of this work was supported by National Science
Foundation grant CHE89-18841. We made use of computer resources
provided under NSF grant CHE88-21737.

\begin{figure}
\caption{Self intermediate scattering function for A particles versus
rescaled time $t/\tau(T)$ (solid curves). Temperatures (from left to
right in the upper part of the figure): 0.466, 0.475, 0.5, 0.55, 0.6,
0.8, 1.0, 2.0, 3.0, 4.0, 5.0.
Dashed curve: von Schweidler law $0.783-0.407(t/\tau)^{0.49}$. Dotted
curve: KWW $0.69\exp(-(t/\tau)^{0.83})$.\label{fig1}}
\vspace*{5mm}
\par
\caption{Self intermediate scattering function for A particles versus
rescaled time $t/\tau(T)$ (solid curves). Temperatures as in Fig.~1.
Dashed curve: von Schweidler law $0.804-0.438(t/\tau)^{0.47}$. Dotted
curve: KWW $0.72\exp(-(t/\tau)^{0.78})$.\label{fig2}}
\vspace*{5mm}
\par
\caption{Intermediate scattering function for AA correlation versus
rescaled time $t/\tau(T)$ (solid curves). Temperatures as in Fig.~1.
Dashed curve: von Schweidler law $0.867-0.475(t/\tau)^{0.52}$. Dotted
curve: KWW $0.81\exp(-(t/\tau)^{0.85})$.\label{fig3}}
\vspace*{5mm}
\par
\end{figure}

\begin{references}
%
\bibitem{bgsleuth84}
U. Bengtzelius, W. G\"otze, and A. Sj\"olander, J. Phys. C {\bf 17}, 5915
(1984); E. Leutheusser, Phys. Rev. A {\bf 29}, 2765 (1984).
%
\bibitem{cummins93}
See e.g. H. Z. Cummins, W. M. Du, M. Fuchs, W. G\"otze, S. Hildebrand,
A. Latz, G. Li, and N. J. Tao, Phys. Rev. E {\bf 47}, 4223 (1993).
%
\bibitem{kob94a}
W. Kob and H. C. Andersen, preprint.
%
\bibitem{kob94b}
W. Kob and H. C. Andersen, (to be published).
%
\bibitem{bibles}
W. G\"otze, p. 287 in {\it Liquids, Freezing and the Glass Transition},
eds. J. P.  Hansen, D. Levesque, and J. Zinn-Justin, Les Houches.
Session LI, 1989, (North-Holland, Amsterdam, 1991); W. G\"otze and L.
Sj\"ogren, Rep. Prog. Phys. {\bf 55}, 241 (1992).
%
\bibitem{hansenmcdonald86}
J.-P. Hansen and I. R. McDonald, {\it Theory of Simple Liquids}
(Academic, London, 1986).
%
\end{references}
\end{document}